# Manifestation of the strong quadrupole light-molecule interaction in the SEHR spectra of symmetrical molecules

**A.M. Polubotko**

A.F. Ioffe Physico-Technical Institute Russian Academy of Sciences, Politechnicheskaya 26, 194021 Saint Petersburg, Russia. Fax: 297-10-17, E-mail: alex.marina@mail.ioffe.ru

The paper demonstrates possibility of giant enhancement of Surface Enhanced Hyper Raman Scattering on the base of qualitative consideration of electromagnetic field near some models of rough metal surfaces and of some features of the dipole and quadrupole light-molecule interaction, such as it was made in the dipole-quadrupole SERS theory. Consideration of symmetrical molecules allows to obtain selection rules for their SEHR spectra and establish such regularity as appearance of the bands, caused by the totally symmetric vibrations, transforming after the unitary irreducible representation in molecules with $C_{nh}, D$ and higher symmetry groups, which are forbidden in usual HRS spectra. Analysis of literature data on trans-1,2-bis (4-pyridyle) ethylene and pyridine molecules demonstrates that their SEHR spectra can be explained by the SEHRS dipole-quadrupole theory, while analysis of the SEHR spectrum of pyrazine reveals appearance of the strong forbidden bands, caused by vibrations transforming after the unitary irreducible representation that strongly confirms this theory which allows to interpret the whole SEHR spectrum in detail. The results corroborate this common mechanism of Surface Enhanced optical processes on molecules adsorbed on rough metal surfaces.

# 1. Introduction

Understanding of mechanism of surface enhancement of optical processes by molecules adsorbed on rough metal surfaces is very important. However in spite of there are some reliable conceptions, which strongly corroborate our point of view, the mechanism is still a matter of debates. There are several theories and approaches, which try to explain these phenomena. The most widespread is the one, which is based on the conception of the enhancement by surface plasmons, which may be excited on a metal surface. One should note that plasmons can not be excited on flat surfaces at the applied incident frequencies. Therefore one usually connects their appearance with existence of surface roughness. The conception of the surface plasmons is well defined on flat metal surfaces [1]. However it becomes indefinite on a random rough surface, which exists in real experiments. In addition (in case we suppose existence of such excitations) the whole field, which affects the molecule is determined as a sum of all surface and other modes, existing near rough surface. Therefore we have a great doubt in the fact that some separate plasmon modes determine the whole giant enhancement.

The second conception is the charge transfer theory, or the charge transfer enhancement mechanism [2] which explains these processes by resonance Raman scattering arising between metal electron states and some unoccupied molecular states, which are named as affinity levels. The main objection against this mechanism is that it occurs only in presence of surface roughness. It is quite ununderstandable, why this mechanism is not observed on single metal surfaces, that was strongly confirmed in numerous experiments of Allan Campion (see [3-6] for example). The second objection is that this mechanism must be observed on some definite frequencies, corresponding to the electron transfer from the electronic metal states to the unoccupied states of the molecule. Then it must occur on the frequencies, which depend on the type of the molecule, while the frequency dependence of the enhancement in SERS is not resonant and SERS is observed practically for all molecules in a very large frequency range on arbitrary rough surfaces [7]. One should note, that the charge transfer enhancement mechanism arose from the experimental fact, that the enhancement in the first layer of adsorbed molecules, which have direct contact with the surface is significantly stronger than the enhancement in the second and upper layers. One should note, that this fact was explained in [8-10]. It was demonstrated, that the first layer effect is caused by a very large difference in the values of the electric field and their derivatives in the vicinity of prominent places with a large curvature just in the first and the second layers of adsorbed molecules and has a pure electrodynamical but not the chemical nature [8-10]. The third and the most important opinion on the nature of the enhancement is the one, that the enhancement is caused by so-called rod effect, or by increase of the electric field near the roughness of the wedge, or (tip or cone, or spike) type [11-17]. The enhancement at the tops of these ideal features is infinity. Therefore this fact guarantees the strong enhancement near the tops of the roughness of the wedge or cone, or spike like forms. This fact well corroborates the main reason of SERS and other surface enhanced processes-the strong surface roughness. In addition this mechanism was well corroborated by experiments of Emory and Nie [18, 19]. However it should be noted, that the pure rod effect is not able to explain such features of the SER, SEHR [7, 20-22] and probably SERR spectra, as appearance of the strong forbidden bands in molecules with $C_{2h}, D$ and higher symmetry. Sometimes this fact is explained by distortion of molecules due to adsorption, however we consider that it is caused by existence of so called strong quadrupole light-molecule interaction and the actual reason of the surface enhanced processes is the strong dipole and especially quadrupole light-molecule interactions arising in surface fields strongly varying in space in the vicinity of the strong surface roughness [10, 17]. Moreover, since we usually deal with random surfaces and random surface electromagnetic fields, the values of amplitudes of separate bands, which can change in a broad range, can not be well defined characteristics, which can well establish the dipole-quadrupole scattering mechanism. The most reliable informative characteristics, which permit to make a unequivocal conclusion about the enhancement mechanism, are the irreducible representations of the vibrations, which cause forbidden and allowed lines and other features in the enhanced spectra of symmetrical molecules. Analysis of the SER spectra of symmetrical molecules [10, 15, 17, 23-26] strongly confirms the dipole-quadrupole mechanism of SERS. Here we shall demonstrate, that the main available regularities of the SEHR spectrum of pyrazine, which refer to the symmetrical molecule with $D_{2h}$ symmetry strongly confirm our point of view and the validity of the dipole-quadrupole enhancement mechanism, arising due to the strong rod effect. In addition the careful analysis of the SEHR spectra of trans-1,2-bis (4-pyridyle) ethylene and pyridine inclines us to opinion that the strong quadrupole light-molecule interaction can manifest in the spectra of the first molecule, while the SEHR spectra of pyridine can be completely explained by the dipole-quadrupole theory.

One should note that our approach, based on such models of the surface roughness as wedge, tip, cone or spike [10-13,15-17] is used in a new branch of enhanced spectroscopy - Tip Enhanced Raman spectroscopy (TERS) [27,28]. The singular behavior near the tops of these models explains the existence of very strong local optical fields associated with the features of the surfaces on the tops of these models, which are responsible for the giant enhancement. One should note, that there can be some another configurations and models of roughness, such as two close parallel round nanowires [29] where the strong enhancement occurs in the gap between them. However in any case the enhancement in our and the other models occurs in some local areas and leads to the enhancement of the electric fields and its derivatives and hence to increase of the dipole and quadrupole light-molecule interactions in these regions.

## 2. Electromagnetic field near rough metal surface

The main property of electromagnetic field near rough metal surface is its strong spatial heterogeneity. As an example we can consider a model of the rough surface - a strongly jagged metal lattice with a regular triangular profile (Figure 1). The electromagnetic field above this lattice can be represented in the form

$$\overline{E} = \overline{E}_{inc} + \overline{E}_{surf..sc} \tag{1}$$

where,

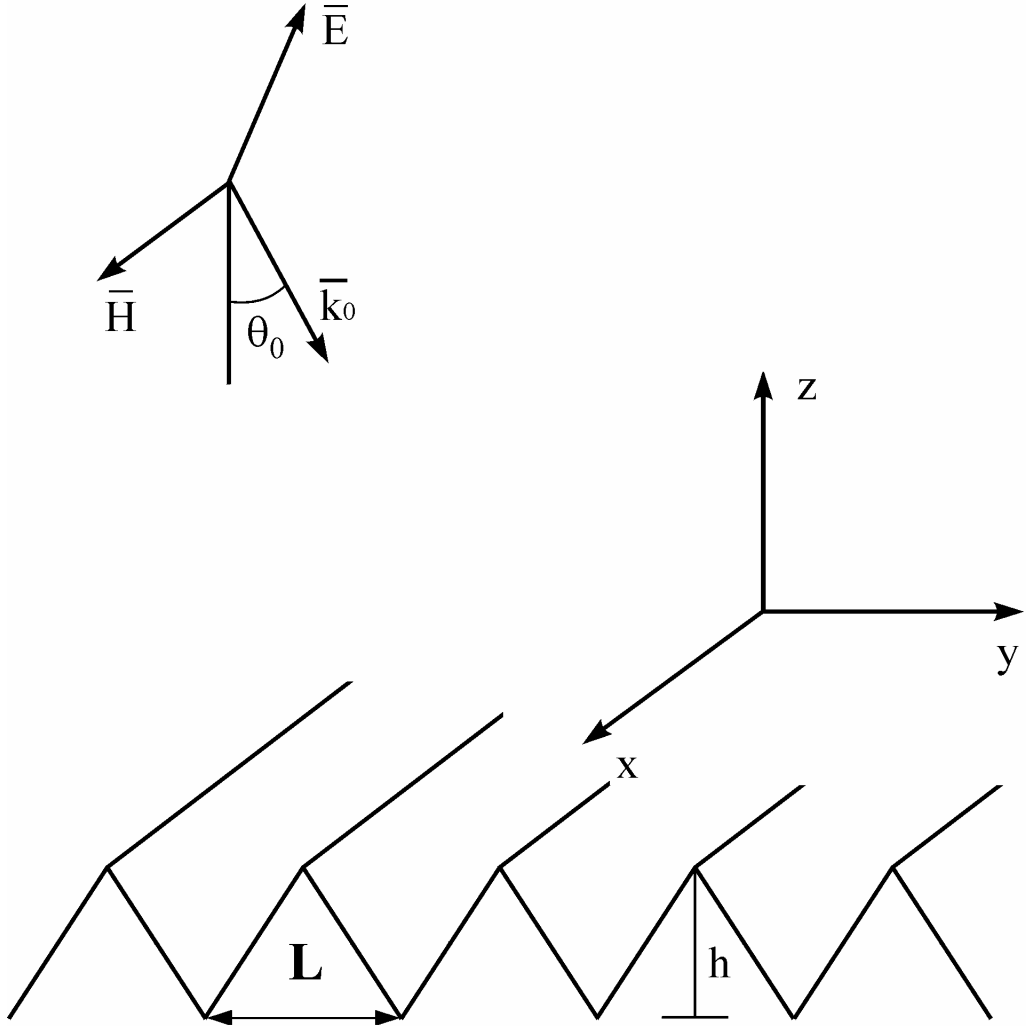

Figure 1. Regular lattice of a triangular profile $(L << \lambda)$. Here $L$ is the period of the lattice and $\lambda$ is the wavelength of the incident light, $h$ is the height of the lattice.

$$\overline{E}_{inc} = \overline{E}_{0,inc} e^{-ik_0 \cos\theta_0 z + ik_0 \sin\theta_0 y}$$

$$\left| \overline{E}_{0,inc} \right| = 1 \tag{2}$$

$\theta_0$ - is the angle of incidence,

$\overline{k}_0$ is the wave vector of the incident field in a free space,

$$\overline{E}_{surf.sc.} = \sum_{n=-\infty}^{+\infty} \overline{g}_n e^{i\alpha_n y + i\gamma_n z} \tag{3}$$

$$\alpha_n = \frac{2\pi n + k_0 \sin\theta_0 L}{L} \tag{4}$$

$$\gamma_n = \sqrt{k_0^2 - \alpha_n^2} \tag{5}$$

Here $\overline{g}_n$ are the amplitudes, $n$ is the number of a spatial harmonic. For the period of the lattice $L << \lambda$ the spatial harmonic with $n = 0$ is a direct reflected wave, while all the others are heterogeneous plane waves strongly localized near the surface. The maximum localization size has the harmonic with $n = 1$. All the others are localized considerably stronger. The exact solution of the diffraction problem on the lattice reduces to determination of the coefficients $\overline{g}_n$. The main specific feature of the surface field is a steep or singular increase of the electric field near the wedges of the lattice or so-called rod effect. This type of behavior is independent on a particular surface profile. It is determined only by existence of sharp wedges. Besides it is independent on the dielectric properties of the lattice and exists in lattices with any dielectric constants that differ from the dielectric constant of vacuum. In the vicinity of the wedge (Figure 2a) the electric field can be estimated as

$$E_r = -g_{0,inc}C_0\left(\frac{l_1}{r}\right)^{\beta}\sin(\lambda_1\varphi)$$

$$E_{\phi} = g_{0,inc}C_0\left(\frac{l_1}{r}\right)^{\beta}\cos(\lambda_1\varphi)$$

(6)

where $C_0$ is some numerical coefficient, $(l_1 = L$ or $h)$ is a characteristic size of the lattice

$$\lambda_1 = \pi/(2\pi - \alpha) \tag{7}$$

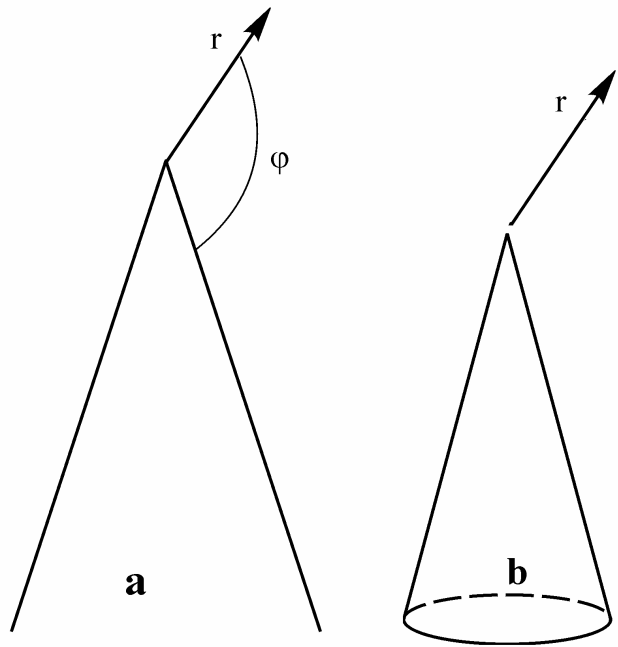

Figure 2. **a**-infinite wedge, **b**-the roughness of the cone type.

$\alpha$ is the wedge angle for an ideally conductive wedge.

$$\beta = 1 - \lambda_1 = \frac{\pi - \alpha}{2\pi - \alpha} \tag{8}$$

The specific feature of the field behavior (6) is appearance of the singularity $(l_1/r)^\beta$, which describes geometrical nature of the field enhancement. It determines the following behavior of the coefficients $\overline{g}_n$ in expression (3) [30]:

$$g_n \sim |n|^{\beta - 1}. \tag{9}$$

Indeed, substitution of (9) into (3) gives

$$\sum_{\substack{n=-\infty \\ n \neq 0}}^{+\infty} |n|^{\beta - 1} e^{2\pi |n| z/L} \sim 2 \int_0^\infty t^{\beta - 1} e^{-2\pi z t/L} dt \sim 2 \left( \frac{L}{2\pi z} \right)^\beta \tag{10}$$

For the wedge angles changing in the interval $0 < \alpha < \pi$ the $\beta$ value varies within the range $0 < \beta < 1/2$ and the coefficients $g_n$ slowly decrease as $n$ increases. Thus the singular behavior of the field arises because of specific summation of the surface waves at the top of the wedge. In the region of a three-dimensional roughness of the cone type (Figure 2b) the formula for estimation of the field has an approximate form

$$E_r \sim g_{0,inc} C_0 \left( \frac{l_1}{r} \right)^\beta, \tag{11}$$

where $\beta$ depends on the cone angle and varies within the interval $0 < \beta < 1$. Using formulae (6) and (11) one can note a very important property: a strong spatial variation of the field. For example

$$\frac{1}{E_r} \frac{\partial E_r}{r} \sim \left( \frac{\beta}{r} \right) \tag{12}$$

can be significantly larger than the value $2\pi/\lambda$, which characterizes variation of the electric field in a free space. If one considers more realistic models of the rough surface than the regular metal lattice, it is obvious, that there is a strong enhancement of the perpendicular component of the electric field at places with a large curvature $\overline{E}_n$ (Figure 3) while the tangential components $\overline{E}_\tau$ are comparable with the amplitude of the incident field. Besides, the electromagnetic field strongly varies in space with a characteristic length $l_E$ equal to characteristic roughness size. This type of behavior is not an exclusive property of the ideally conductive lattice and preserves near surfaces with a finite dielectric constant.

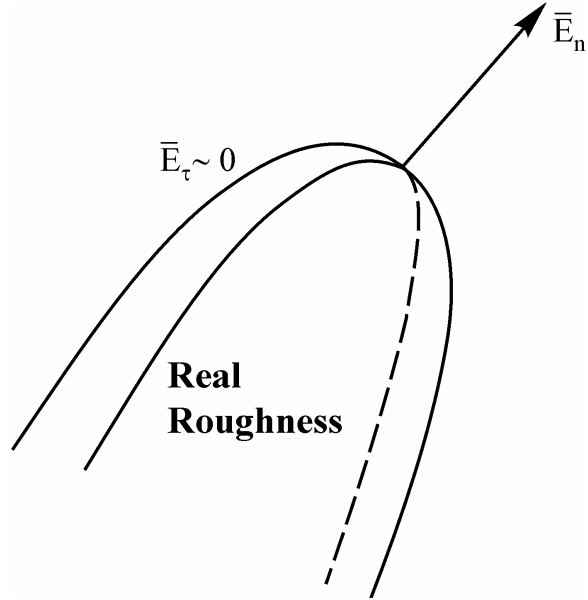

Figure 3. More realistic model of the roughness of the cone type. The normal component of the electric field $\overline{E}_n$ and the derivatives $\partial E_\alpha / \partial x_\alpha, \alpha = (x, y, z)$ are enhanced near the top of the roughness.

One should note that there is a number of works, which try to calculate the main features of the surface electromagnetic field near rough surface by numerical methods for various models of roughness. They are the discrete dipole approximation DDA method [31], the finite difference time-domain method (FDTD) [32, 33], the method of volume integral equation [34] and may be some others. The features of these methods one can find in the above references. It should be noted, that an electromagnetic field is a long-range field. Therefore the electric field in a definite point depends on the field values in a large area around the point. However one of the deficiency of the DDA and FDTD methods is limitation in a number of discrete elements, which are involved in calculations and hence limitation of the size of the area used in calculations, which can result in definite uncertainty in the obtained results. Therefore in spite of the above methods may consider as a more precise methods than the our one, they can give large uncertainties in the values of the calculated fields. However appearance of novel power computers can essentially improve this situation due to possibility of increasing of the number of the elements. Below we shall describe some results, obtained by these methods for some models of roughness and will compare them with the results obtained by our approach. It appears that the obtained results confirm strongly our point of view on the behavior of the field. In [32] the authors calculated the enhancement near some conical tip with round tip apex. The results qualitatively coincide with our ones. In fact there is a strong enhancement of the electric field in a narrow area of the round tip apex. In spite of the authors name this behavior as a manifestation of tip plasmons, this result do not differ from the behavior of the electric field due to the rod effect. Another series of works are the works of Kottmann et. al. [34-36] who calculated the behavior of the electric field near some models of nanowires using the numerical method of the volume integral equation [34]. In spite of the authors assign the enhancement to plasmon resonances, they measured the enhancement in the vicinity (1nm) of the wedges of hexagon, pentagon, square and triangle nanowires (Figure 4), where the rod effect is very strong.

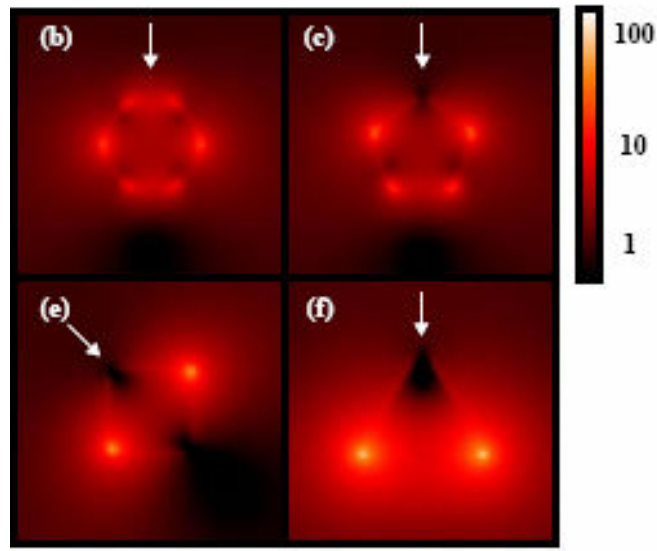

Figure 4. Diffraction of a plane electromagnetic wave on nanowires of hexagon, pentagon, square and triangle forms. The wave vector is perpendicular to the main axis of nanowires, while the $\overline{E}$ vector is in the plane of the figure [36]. One can see the strongest enhancement of the electric field in the vicinity of the wedges of nanowires that is manifestation of the rod effect.

The measured distance dependence of the enhancement from the tops of the wedges (Figure 5)

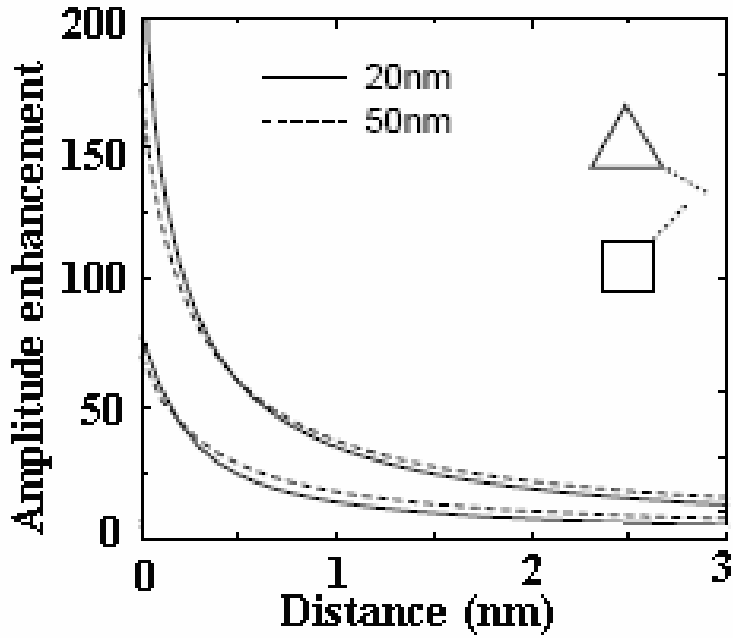

Figure 5. The distance dependences of the enhancement of the electric field from the wedges of nanowires of the square and triangle forms of 20 and 50 nm sizes [36]. The dependences qualitatively correspond to the behavior, described by formulae (6,11).

qualitatively confirms the dependences, described by formulae (6, 11). Because of the difference of the models in [35,36], and specificity of our and real models of rough surface, the calculated wavelength dependences of the enhancement (Figure 6 for example) does not reflect the main features of experimental wavelength dependences of SERS for arbitrary rough surfaces [7] (Figure 7 for example).

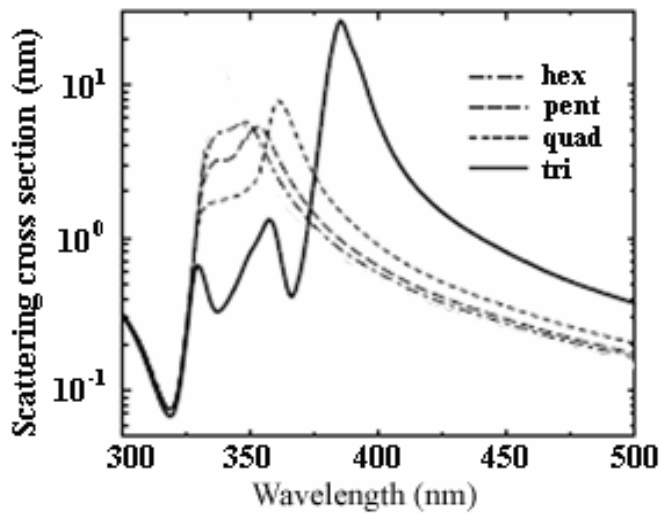

Figure 6. The wavelength dependences of the enhancement of the electric field near the wedges of nanowires of hexagon, pentagon, square and triangle forms [36]. One can see that the region of resonances corresponds to the wavelengths less than ~425nm.

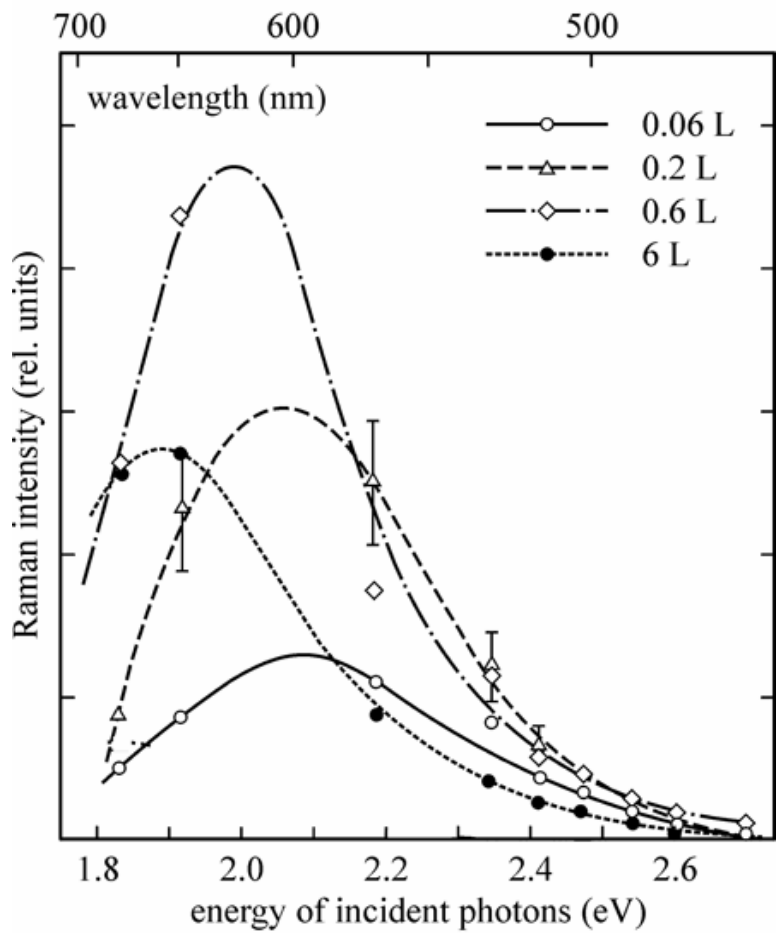

Figure 7. Typical wavelength dependences of the intensity of the SER bands for real rough surface of silver [7]. One can see that the region of the main enhancement corresponds to the wavelengths of the incident radiation $\lambda \geq 450$ nm.

The region of wavelengths of the resonances, calculated in [35, 36] is approximately (300-425) nm (Figure 6) with significantly smaller wavelengths than the region of the most enhancement in SERS on a real rough surface (450-700) nm (Figure 7). Taking into account, that the enhancement, caused by the nanowires can be measured at smaller distances from the wedges than 1 nm (as it was made in the calculations), the enhancement of the electric field can be significantly stronger than those obtained in [35,36]. Therefore all these results confirm the fact that the strongest enhancement, which arises in the vicinity of the wedges of the nanowires in the region of wavelengths, where the strong SERS is observed is caused not by the plasmon resonances but by the rod effect. One should note, that the detailed consideration of the derivatives of the electric field is absent in these papers such as in [32] and the real enhancement factor, which further is considered in our work is not considered in [35, 36].

As it was mentioned above there is calculation of the electric field between two close round nanowires, with parallel axis's [29]. It was established a strong enhancement of the electric field in the gap between them (Figure 8)

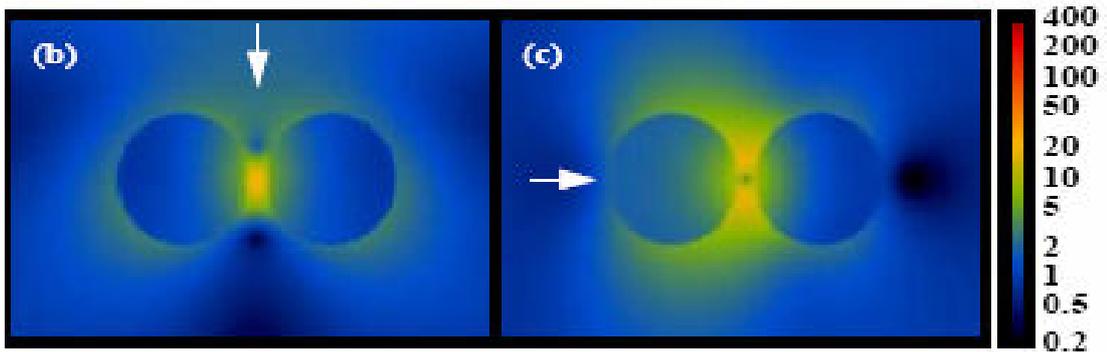

Figure 8. Diffraction of a plane electromagnetic wave on two coupled cylindrical nanowires of 50 nm. The incident direction is designated by arrows. The electric field is in the plane of the figure. One can see a region of the enhanced electric field in the gap between the nanowires [29].

and some resonances in the wavelength region (325-425) nm (Figure 9).

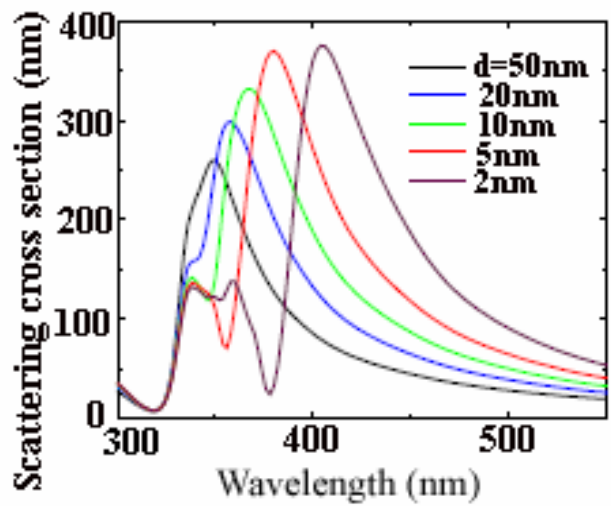

Figure 9. The wavelength dependence of the scattering cross-section for 50 nm coupled nanowires for various distances d between them [29]. One can see that the region of resonances for these models corresponds to the wavelengths $\lambda \leq$ (425-450) nm.

However these resonances are situated in another wavelength range, than the range for maximum SERS (450-700) nm as in the previous case (Figure 7). Therefore this model can not explain the enhancement on arbitrary rough surface by these resonances too. This result is important from our point of view since it points out the existence of some additional active areas, or active sites (in A. Otto terminology [2]) where the strong enhancement of the electric field is possible. Analogous calculations by the finite-time domen method (FTDT) were presented in [37] for some chains of dimers. It was established that such areas of the enhancement of the electric field (active sites) or hot spots (in terminology of V. M. Shalaev) exist between two spheres of dimers too. However the strongest enhancement arises for the dimer consisting of tetrahedrons. The field enhancement is ~ $5.62 \times 10^2$ and ~$10^{11}$ for SERS for single dimer of two truncated tetrahedrons with 1 nm gap between their tops. These values can be increased in the lattice of dimers, using the long range effects, or space resonances, arising in their periodical chains. The field enhancement can achieve $1.77 \times 10^3$, while the enhancement for SERS ~$10^{13}$. The essential point of these investigations is that the maximum enhancement arises in the dimers with truncated tetrahedrons. Thus we again dealt with the rod effect in the vicinity of the truncated tops of tetrahedrons and the main reason of this enhancement is the strong increase of the electric field near their tops, which are a good model of sharp roughness.

The ideal enhancement of the field at the top of the wedge, cone, tip or spike is infinity. The same result must be for the ideal sharp tops of tetrahedrons. However the existence of the truncation decreases strongly the SERS enhancement till $10^{13}$. One should note, that we achieved the SERS enhancement ~ $1.7 \times 10^{16}$ in our estimations for Single Molecule SERS [38] for the quadrupole enhancement mechanism. However one must take into account, that in all considered models we have some arbitrary parameters, which are defined from some reasonable physical point of view. Thus in spite of we use some another models of the roughness and their parameters than the authors of [37] the reason of the enhancement in all considered models is the same-the rod effect arising near sharp wedges or tips, or prominent places with a large curvature.

Thus the second type of the active sites, or hot spots arise in the gap between two parallel round nanowires or two spherical particles. The large enhancement can arise in the gap between two tops of tetrahedrons, cones, tips or spikes. The last cases are some limited cases of the existence of the enhancement in the gap between two nanowires [29], or between two spheres [37]. Moreover the enhancement in the last cases apparently can be the largest for sufficiently thin spikes with a small angle of the spike apex and a very small gap between them. In addition it must depend and be limited by their length, which must be equal or more than the wavelength of the incident field for the maximum enhancement. The last result follows from consideration of wave and electrostatic approximation. The analog of this enhancement geometry is two close spikes, situated on the same axis with a small gap between tops of the spikes (Figure 10).

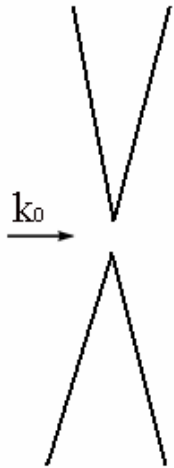

Figure 10. The limited case of the active sites of the second type. Apparently the most enhancement can be obtained in this geometry.

In this case they create electric arc between their tops under influence of the incident field. Apparently this geometry possesses by maximum enhancement. One must take into account that we have some arbitrary parameters which are defined from some reasonable physical point of view in the all considered models. Thus in spite of we use some another models of the roughness and their parameters than the authors of [35-37] the reason of the enhancement in the all considered models is the same-the rod effect arising near sharp wedges, tips, cones or spikes, or prominent places with a large curvature, or the field arising in the gap between tops of the above features.

Since for further consideration we use the main property of these models-the increase of the electric fields and their derivatives near the above features and qualitative consideration of the enhancement of the dipole and quadrupole light molecule interactions, the choice between these models is not of principle. Therefore further we shall use our models of the wedge or cone (spike), and formulae (6, 11) for consideration of the principal enhancement in SERS and SEHRS since it is sufficient for our goals.

## 3. Interaction of light with molecules near rough metal surface

In accordance with principles of theoretical physics the optical properties of molecules are determined by the light-molecule interaction Hamiltonian, which has the form

$$\hat{H}_{e-r} = -\sum_i \frac{ie\hbar}{mc} \overline{A}_i \overline{\nabla}_i \qquad (13)$$

Here the sign of $e$ is positive. $\overline{A}_i$ is a vector potential of the electromagnetic field at the place of the $i$ electron. Other designations are conventional. For small objects, like molecules, the vector potential can be expanded in a Taylor series and final expressions for the light-molecule interaction Hamiltonians for the incident and scattered fields can be obtained in the form

$$\hat{H}_{e-r}^{inc} = \left| \overline{E}_{inc} \right| \frac{(\overline{e}^* \overline{f}_e^*)_{inc} e^{i\omega_{inc}t} + (\overline{e} \overline{f}_e)_{inc} e^{-i\omega_{inc}t}}{2} \qquad (14)$$

$$\hat{H}_{e-r}^{scat} = \left| \overline{E}_{scat} \right| \frac{(\overline{e}^* \overline{f}_e^*)_{scat} e^{i\omega_{scat}t} + (\overline{e} \overline{f}_e)_{scat} e^{-i\omega_{scat}t}}{2} \qquad (15)$$

where $\overline{E}_{inc}$ and $\overline{E}_{scat}$ are vectors of the incident and scattered electric fields, $\omega_{inc}$ and $\omega_{scat}$ are corresponding frequencies, $\overline{e}$ -are polarization vectors of the corresponding fields,

$$f_{e\alpha} = d_{e\alpha} + \frac{1}{2E_\alpha} \sum_\beta \frac{\partial E_\alpha}{\partial x_\beta} Q_{e\alpha\beta} \qquad (16)$$

is an $\alpha$ component of the generalized vector of interaction of light with molecule,

$$d_{e\alpha} = \sum_i ex_{i\alpha} , \quad Q_{e\alpha\beta} = \sum_i ex_{i\alpha}x_{i\beta} \qquad (17)$$

are the $\alpha$ component of the dipole moment vector and the $\alpha\beta$ component of the quadrupole moments tensor of interaction of light with electrons of the molecule. Here under $x_{i\alpha}$ and $x_{i\beta}$ we mean coordinates $x, y, z$ of the $i$ electron. Usually the relative influence of the quadrupole and dipole light-molecule interactions is determined as the relation of the second and the first terms in the right side of (16). However this relation is the relation of quantum-mechanical operators, while all physical values are expressed via the matrix elements of these operators. Therefore one must consider, that the relative influence is determined by the relations

$$\frac{\overline{\langle m|Q_{e\alpha\beta}|n\rangle}}{\overline{\langle m|d_{e\alpha}|n\rangle}} \frac{1}{2E_\alpha} \frac{\partial E_\alpha}{\partial x_\beta} = B_{\alpha\beta} a \frac{1}{2E_\alpha} \frac{\partial E_\alpha}{\partial x_\beta} \qquad (18)$$

where $a$ is a molecule size, $B_{\alpha\beta}$ are some numerical coefficients. The first factor in the left side of (18) is the relation of some mean matrix elements of the quadrupole and dipole transitions. It is necessary to point out that the $B_{\alpha\beta}$ values essentially differ for $\alpha \neq \beta$ and for $\alpha = \beta$. This fact results from the fact that $d_{e\alpha}$ and $Q_{e\alpha\beta}$ are the values with a changeable sign, while $Q_{e\alpha\alpha}$ are the values with a constant sign, that strongly increases the $B_{\alpha\alpha}$ values. The difficulty for this estimation is the fact, that there is no information about quadrupole transitions in molecules. Therefore because the inner shell configuration in molecules remains almost intact we usually take the value $\overline{\langle n|Q_{e\alpha\alpha}|n\rangle}$ instead of $\overline{\langle m|Q_{e\alpha\alpha}|n\rangle}$ and the value $\sqrt{e^2\hbar/2m\omega_{mn}} \times \sqrt{\overline{f}}_{mn}$ for $\overline{\langle m|d_\alpha|n\rangle}$, which is expressed in terms of some mean value of the oscillator strength $\overline{f}_{mn} = 0.1$ while $\omega_{mn}$ corresponds to the edge of absorption ($\lambda \approx 2500A$). Since configuration of the electron shell is close to configuration of nuclei the value $\overline{\langle n|Q_{\alpha\alpha}|n\rangle}$ was calculated as a $Q_{n\alpha\alpha}$ component of the quadrupole moments of nuclei. Estimation for the pyridine, benzene or pyrazine molecules gives the value $B_{\alpha\alpha} \sim 2\times10^2$ that strongly differs from the value $B_{\alpha\alpha} \sim 1$ ($B_{\alpha\alpha}a \sim a$). The last one is usually used in literature as the relation of the quadrupole and dipole operators. The full estimation of the relation (18) near rough metal surface is made for the models of roughness of the wedge or the cone or the spike types, which well approximate prominent features of the surface with a large curvature. Then using (12) one can obtain the following expression for

$$\frac{\overline{\langle m|Q_{e\alpha\alpha}|n\rangle}}{\langle m|d_{e\alpha}|n\rangle}\frac{1}{2E_\alpha}\frac{\partial E_\alpha}{\partial x_\alpha} = B_{\alpha\alpha}a\frac{\beta}{r} \tag{19}$$

It can be seen, that for $r < B_{\alpha\alpha}a\beta$ the quadrupole interaction can be more than the dipole one. The enhancement of the electric field and the dipole interaction can be estimated as

$$G_{H_d} \sim C_0\left(\frac{l_1}{r}\right)^\beta \tag{20}$$

while the enhancement of the quadrupole interaction compared to the dipole interaction in a free space can be estimated as

$$G_{H_Q} \sim C_0\beta\left(\frac{B_{\alpha\alpha}}{2}\right)\left(\frac{l_1}{r}\right)^\beta\left(\frac{a}{r}\right) \tag{21}$$

It can be seen, that for reasonable values $C_0 \sim 1$, $l_1 \sim 10\,\text{nm}$, $r \sim 1\,\text{nm}$, $\beta \sim 1$ and the molecules like pyridine, benzene or pyrazine with $B_{\alpha\alpha} \sim 2\times 10^2$ the enhancement of the dipole interaction is ~10, while the enhancement of the quadrupole interaction is $\sim 10^2$.

## 4. The enhancement in SEHRS

The enhancement in SEHRS such as in SERS can be caused both by the enhancement of the $E_z$ component of the electric field, which is perpendicular to the surface and by the enhancement of the field derivatives and the quadrupole interaction. Since SEHRS is the process of the third order the enhancement due to the dipole interaction is

$$G_d \sim C_0^6\left(\frac{l_1}{r}\right)^{6\beta} \tag{22}$$

and

$$G_Q \sim C_0^6 \beta^6 \left(\frac{B_{\alpha\alpha}}{2}\right)^6 \left(\frac{l_1}{r}\right)^{6\beta} \left(\frac{a}{r}\right)^6 \qquad (23)$$

due to the quadrupole interaction. For the values of the parameters pointed out above, the enhancement due to the purely dipole interaction is of the order of $10^6$ while the enhancement due to the quadrupole interaction $\sim 10^{12}$. One should note that the enhancement соеаашсшутеы of the dipole and quadrupole interactions $G_{H_d}$ and $G_{H_Q}$ and especially the enhancement of SEHRS $G_Q$ can be very large. For example for some limited situations with the values of the parameters $C_0 \sim 1$, $B_{\alpha\alpha} \sim 2 \times 10^2$, $r \sim 0.1 nm$, $\beta \sim 1$, $l_1 \sim 100 nm$ corresponding to the placement of the molecule on the top of the cone (tip or spike) the enhancement сщуаашсшуте $G_Q$ in SEHRS can achieve $10^{30}$. As it was mentioned above the real situation is that most enhancement arises in the vicinity of some points associated with prominent places with very large curvature. The mean enhancement is formed from the whole layer of adsorbed molecules and is significantly smaller than the maximum enhancement near these places due to averaging.

## 5. Main and minor moments

In accordance with the previous consideration the most enhancement arises from the quadrupole interaction with the moments $Q_{xx}, Q_{yy}, Q_{zz}$ and the dipole interaction caused by the enhancement of the electric field, which is perpendicular to the surface. Let us designate the coordinate system, associated with the molecule as $(x, y, z)$, while the coordinate system associated with the surface as $(x', y', z')$ with the $z'$ axis perpendicular to the surface. Depending on orientation of the molecule the dipole light molecule interaction with various components of the $d$ moments can be important for the enhancement. For example we shall consider that the pyrazine molecule (which will be considered below) is situated in the $XZ$ plane when the $z$ axis passes through both nitrogen atoms. Then for the pyrazine molecule lying flatly on the surface (with the $d_y$ moment perpendicular to the plane of the molecule) only the $d_y$ moment, which is parallel to the $E_{z'}$ component of the enhanced electric field is essential. For the molecule bounded by the nitrogen atom with the surface the essential $d$ moment is $d_z$. Usually all experiments are performed with molecules in solutions. The most enhancement arises from molecules, which are closest to the surface (the first layer effect). Then the molecules can be arbitrary oriented with respect to the surface and to the $E_{z'}$ component of the electric field and all the $d$ moments can contribute to the scattering. Further we shall name all the essential moments as main moments, while the nonessential moments $Q_{xy}, Q_{xz}, Q_{yz}$ as the minor ones.

Since our further consideration concerns symmetrical molecules and SEHRS selection rules in these molecules, it is necessary to determine the minor and the main moments for this case. It is convenient to transfer to the values, which transform after irreducible representations of the symmetry group. Analysis of the tables of irreducible representations of all point groups [10, 39] demonstrates that the $d$ and $Q_{xy}, Q_{xz}, Q_{yz}$ moments transform after irreducible representations in the most part of the point symmetry groups, while the $Q_{xx}, Q_{yy}, Q_{zz}$ moments can be expressed via linear combinations $Q_1, Q_2, Q_3$ transforming after irreducible representations. (Further we shall consider only such molecules. However the results obtained here can be transferred on molecules with any point group).

$$Q_{xx} = a_{11}Q_1 + a_{12}Q_2 + a_{13}Q_3$$
$$Q_{yy} = a_{21}Q_1 + a_{22}Q_2 + a_{23}Q_3 \qquad (24)$$
$$Q_{zz} = a_{31}Q_1 + a_{32}Q_2 + a_{33}Q_3$$

where the coefficients $a_{ij}$ depend on the symmetry group. The corresponding expressions for $Q_1, Q_2, Q_3$ are

$$Q_1 = b_{11}Q_{xx} + b_{12}Q_{yy} + b_{13}Q_{zz}$$
$$Q_2 = b_{21}Q_{xx} + b_{22}Q_{yy} + b_{23}Q_{zz} \qquad (25)$$
$$Q_3 = b_{31}Q_{xx} + b_{32}Q_{yy} + b_{33}Q_{zz}$$

Here the coefficients $b_{ij}$ depend on the symmetry group too. The specific form of $Q_1, Q_2, Q_3$ for some point groups one can find in [10] and in Appendix. There are combinations with a constant sign, which can be named as the main moments and of a changeable sign, which can be named as the minor ones. In accordance with our previous consideration the main moments are responsible for the strong enhancement, while the minor moments are nonessential for the scattering. For trans-1,2-bis (4-pyridyle) ethylene, pyridine and pyrazine molecules, which are considered in this paper $Q_1, Q_2$ and $Q_3$ coincide with $Q_{xx}, Q_{yy}$ and $Q_{zz}$

## 6. SEHRS in symmetrical molecules

The above estimations give only understanding of the principle possibility of the SEHRS enhancement mechanism. A good prove of its validity can be obtained from consideration of selection rules and regularities of the SEHR spectra of symmetrical molecules. The detailed mathematical consideration of SEHRS is very cumbersome. However one can receive main results from some physical point of view on the base of the methods used in the SERS theory [10]. SEHRS is a three photon process, which occurs via the dipole and quadrupole moments. In case we consider the wavefunctions of symmetrical molecules transforming after irreducible representations of the molecule symmetry group, the full cross-section of the $(s, p)$ vibrational band can be expressed as a sum of various scattering contributions, which occur via various combinations of the dipole and quadrupole moments such as in SERS (formula 59 in [10]).

$$\sigma_{SEHRS(s,p)} \sim \left| \sum_{f_1, f_2, f_3} T_{(s,p), f_1 - f_2 - f_3} \right|^2 \tag{26}$$

Here $f_1, f_2, f_3$ are various dipole and quadrupole moments, transforming after irreducible representations of the symmetry group of the molecule. The values of contributions $T_{(s,p), f_1 - f_2 - f_3}$ are expressed via the values

$$R_{n,l,(s,p)} \langle n | f_1 | m \rangle \langle m | f_2 | k \rangle \langle k | f_3 | l \rangle \tag{27}$$

and similar expressions, with various permutations of the moments. Here $(s, p)$ designates the quantum numbers of degenerated vibrations of the molecule. $s$ numerates the group of degenerated vibrations, while $p$ numerates the vibrations inside the group. $n$ designates the ground, while $l, k, m$ the excited states of the molecule. $R_{n,l,(s,p)}$ are the coefficients of excitation of the $l$ electronic states due to vibration $(s, p)$. Their evident form can be found in [10, 17, 24]. In accordance with consideration, similar to [10, 17, 24], the following selection rules can be obtained

$$\Gamma_{(s,p)} \in \Gamma_{f_1} \Gamma_{f_2} \Gamma_{f_3} \qquad (28)$$

where the symbol $\Gamma$ designates the irreducible representation of the corresponding moment $f$ and of the vibration $(s, p)$. The physical sense of the expressions of the type (27) is that the contributions are expressed via the sequence of quantum transitions arising via various dipole and quadrupole moments and their permutations. In accordance with our previous consideration the most enhancement for the strongly rough surface is caused by the quadrupole interaction with $Q_{main}$ moments and by the dipole interaction with the dipole moments $d_{main}$, which have a non zero projection to the $E_{z'}$ component of the electric field which is perpendicular to the surface. Then the contributions $T_{(s,p), f_1 - f_2 - f_3}$ which we shall designate further simply as $(f_1 - f_2 - f_3)$ can be classified qualitatively after the enhancement degree in the following manner:

1. $(Q_{main} - Q_{main} - Q_{main})$ - the most enhanced scattering type.

2. $(Q_{main} - Q_{main} - d_{main})$ - scattering type, which can be strongly enhanced too, but in a lesser degree than the previous one,

3. $(Q_{main} - d_{main} - d_{main})$ - scattering type, which can be strongly enhanced too, but lesser, than the two previous ones and

4. $(d_{main} - d_{main} - d_{main})$ - scattering type, which can be strongly enhanced too, but lesser than the three previous ones.

Here and further we mean under $(f_1 - f_2 - f_3)$ all contributions with permutations of the $f$ moments. Considering molecules with $C_{nh}, D$ and higher symmetry one can note, that the first and the third enhancement types, with the same $d_{main}$ moments contribute to the bands, caused by vibrations transforming after the unitary irreducible representation or by the totally symmetric vibrations, the second types contribute to the bands, caused by vibrations transforming such as the $d_{main}$ moment, while the forth types also can be significantly enhanced and can contribute to the bands with various symmetry types. Thus the most enhanced bands in the molecules with $C_{nh}, D$ and higher symmetry are caused by the above types of vibrations. The first ones are forbidden in usual HRS. Thus these bands must be the essential feature of the SEHR spectra of symmetrical molecules with the above symmetry. The other contributions without $Q_{main}$, or those, which contain $Q_{\min or}$ moments apparently will be small.

## 7. Analysis of the SEHR spectra of trans-1, 2-bis (4-pyridyle) ethylene

The amount of the works, which analyze the SEHR spectra of symmetrical molecules is very small, such as the number of works on other aspects of SEHRS. Here and further we shall analyze some previous results and peculiarities of the SEHR spectra of, trans-1, 2-bis (4-pyridyle) ethylene [40] pyridine [41, 42, 21, 22] and pyrazine [21, 22], published and interpreted by another authors. The main feature of all these works is consideration of the SEHR spectra using the dipole approximation of the light-molecule interaction Hamiltonian only. In addition the authors of [40-42, 22] use very approximate numerical methods for calculations of various characteristics of Hyper Raman processes like vibration wavenumbers (frequencies), polarizabilities, hyperpolarizabilities, their normal coordinate derivatives, band intensities and some others. From our point of view the neglecting by the strong quadrupole light molecule interaction and the use of these methods result in a very large discrepancy of the experimental results on these spectra and their calculating values. Let us consider the results, which refer to trans-1, 2-bis (4-pyridyle) ethylene in the adsorption geometry, when the long $y$ axis of the molecule is perpendicular to the surface. In accordance with the above SEHRS theory only the bands of $A_u$ and $B_u$ symmetry can be observed in the dipole approximation. This result corresponds to the results and conditions published in [40]. However consideration of the calculated vibration wavenumbers of the bands with the $A_g$ and $B_u$ symmetry demonstrates nearly the same values for major vibration wavenumbers within the calculation errors (0-10) $cm^{-1}$ (Table 1). In this situation assignment of the experimental SEHR bands to $A_g$ and $B_u$ symmetry types is impossible, since the uncertainty in determination of the calculated wavenumbers of the bands with this symmetry is less then the uncertainty of assignment of the measured to the calculated values, which can differ one from another till (50-60) $cm^{-1}$. Thus in accordance with the results of Table 1 many experimental SEHR bands, which were assigned to the $B_u$ irreducible representation can be assigned to the $A_g$ symmetry type. Moreover comparison of the wavenumbers of the bands with $B_u$ symmetry measured in SEHRS experiments with the ones of $A_g$ symmetry measured in SERS experiments demonstrates a very small difference ~ (1-10) $cm^{-1}$ for a large number of the bands. Thus both results incline us to opinion that many SEHRS bands of $B_u$ symmetry can be assigned in fact to $A_g$ symmetry. Taking into account, that the bands with $A_g$ symmetry are allowed in the SEHR spectra of 1, 2-bis (4-pyridyle) ethylene in the dipole-quadrupole theory, the above facts do not contradict to explanation of this SEHR spectra in terms of the strong dipole and quadrupole light-molecule interactions.

Table 1. Calculated and experimental wavenumbers of the SER and SEHR bands of trans-1, 2-bis (4-pyridyle) ethylene and their symmetry types in accordance with [40].

| The numbers of vibrations of $A_g$ symmetry | Calculated wavenumbers of $A_g$ symmetry $(cm^{-1})$ | The numbers of vibrations of $B_u$ symmetry | Calculated wavenumbers of $B_u$ symmetry $(cm^{-1})$ | Experimental wavenumbers for SEHRS of $B_u$ symmetry $(cm^{-1})$ | Experimental wavenumbers for SERS of $A_g$ symmetry $(cm^{-1})$ |
|---|---|---|---|---|---|
| 17 | 280 | 16 | 466 | 552 | 320 |
| 16 | 645 | 15 | 537 | 599 | 652 |
| 15 | 675 | 14 | 677 | 688 | 663 |
| 14 | 866 | 13 | 814 | 841 | 847 |
| 13 | 986 | 12 | 987 | 972 | 1008 |
| 12 | 1071 | 11 | 1071 | 1007 | 1064 |
| 11 | 1097 | 10 | 1095 | 1116 | |
| 10 | 1145 | 9 | 1137 | 1198 | 1200 |
| 9 | 1201 | 8 | 1221 | 1208 | 1200 |
| 8 | 1224 | 7 | 1239 | 1289 | 1244 |
| 7 | 1340 | 6 | 1302 | 1325 | 1314 |
| 6 | 1363 | 5 | 1366 | 1341 | 1338 |
| 5 | 1409 | 4 | 1418 | 1422 | 1421 |
| 4 | 1498 | 3 | 1504 | 1489 | 1493 |
| 3 | 1560 | 2 | 1569 | 1548 | 1544 |
| 2 | 1609 | 1 | 1610 | 1593 | 1604 |
| 1 | 1682 | | | | 1640 |
| | 3018 | | 3012 | | |
| | 3034 | | 3033 | | |
| | 3049 | | 3049 | | |
| | 3061 | | 3061 | | |
| | 3069 | | 3069 | | |
| The numbers of vibrations of $B_g$ symmetry | Calculated wavenumbers of $B_g$ symmetry $(cm^{-1})$ | The numbers of vibrations of $A_u$ symmetry | Calculated wavenumbers of $A_u$ symmetry $(cm^{-1})$ | Experimental wavenumbers for SEHRS of $A_u$ symmetry $(cm^{-1})$ | Experimental wavenumbers for SERS of $B_g$ symmetry $(cm^{-1})$ |
| 8 | 409 | 9 | 298 | 306 | 400 |
| 7 | 503 | 8 | 408 | 385 | 491 |
| 6 | 747 | 7 | 572 | 664 | 738 |
| 5 | 829 | 6 | 761 | 803 | 802 |
| 4 | 900 | 5 | 865 | 878 | 881 |
| 3 | 956 | 4 | 901 | | 955,972 |
| 2 | 1029 | 3 | 1004 | | 1064 |
| 1 | 1047 | 2 | 1033 | 1060 | |
| | | 1 | 1047 | | |

First we must consider real possible orientations of the trans-1, 2-bis (4-pyridyle) ethylene molecules with respect to the surface. In principle the molecule can adsorb in two main manners-vertically with the main $y$ axis perpendicular to the surface and in flat orientation. Since interaction of the molecule with the surface is weak, one can consider, that the molecule symmetry preserves for both orientations [10, 17, and 24]. The enhancement of the bands of $A_g$ symmetry is caused mainly by $(Q_{main} - Q_{main} - Q_{main})$, $(Q_{main} - d_z - d_z)$ contributions in the horizontal orientation and by $(Q_{main} - Q_{main} - Q_{main})$ and $(Q_{main} - d_y - d_y)$ in the vertical orientation in accordance with classification of the moments after enhancement degree and the selection rules (28). In fact these lines must exist for both orientations. The enhancement of the bands with $A_u$ symmetry is caused mainly by the $(Q_{main} - Q_{main} - d_z)$ and $(d_z - d_z - d_z)$ contributions in the horizontal orientation and by $(Q_{main} - Q_{main} - d_z)$ and $(d_y - d_y - d_z)$ in the vertical orientation, where $d_y$ is perpendicular to the surface and parallel to the $E_z$ component of the electric field. The last contributions apparently are small, because they are caused by $d_z$ moment and by $E_y$ tangential component of the electric field which is not enhanced. The bands with $B_u$ symmetry are caused mainly by $(Q_{main} - Q_{main} - d_{\min or})$ and $(d_z - d_z - d_{\min or})$ scattering types, where under $d_{\min or}$ we mean $d_x$ and $d_y$ components of the dipole moment in the horizontal orientation and mainly by $(Q_{main} - Q_{main} - d_y)$ and $(d_y - d_y - d_y)$ contributions, which can be strongly enhanced in the vertical orientation. In principle the bands with $B_g$ symmetry can appear in the SEHRS spectra too mainly due to existence of the contributions $(Q_{main} - d_z - d_x)$ and $(Q_{main} - d_z - d_y)$ in the horizontal orientation and mainly due to $(Q_{main} - d_y - d_z)$ contributions in the vertical orientation. Analysis of experimental SERS and SEHRS bands demonstrates existence of very close values of wavenumbers of the bands with $A_g$ and $B_g$ symmetry in SERS and of $B_u$ and $A_u$ symmetry in SEHRS (Table 1). Taking into account the large uncertainty in determination of symmetry of the SERS and SEHRS bands due to the calculation errors, one can make a conclusion, that a part of the SEHRS lines can belong to $A_g$ and $B_g$ symmetry. Regretfully we are not able to explain the SEHRS results of [40] more precisely because of the large negligence in calculations of the wavenumbers and in determination of the bands symmetry in this work. However consideration of the strong quadrupole light-molecule interaction and of the two possible orientations of the molecules allows to explain actually the possibility of appearance of the bands with $A_g$ and $B_g$ symmetry in the SEHR spectra of trans-1, 2-bis (4-pyridyle) ethylene.

# 8. Analysis of the SEHR spectrum of pyridine

The SEHR spectra of pyridine were presented in [21, 22, 41, 42]. In principle the spectra differ slightly one from another, that can be explained by various experimental conditions. However the positions of the observed SEHRS lines are nearly the same for all these spectra. Therefore we shall analyze the spectrum, obtained in [41] (Figure 11). The dipole approximation is able to explain main peculiarities of the SEHR spectrum of pyridine. This fact is associated with specific symmetry of this molecule and with the fact, that the $d_z$ moment of the molecule transforms in accordance with the unitary irreducible representation of the $C_{2v}$ symmetry group. It is the reason that all the observed lines are allowed in the dipole approximation. Therefore pyridine is not the molecule which succeeds in discovery of the strong quadrupole light-molecule interaction. However we can reconsider the SEHRS results of [41, 42] on the base of the dipole-quadrupole theory and explain all features of the SEHR spectra. First one should note that the calculated wavenumbers of the pyridine vibrations, obtained in [41] differ strongly from the values, observed experimentally from the SEHR spectrum. This result apparently associated with the use of very approximate methods of determination of vibration numbers in [41]. However the experimental wavenumbers well fit the values, published in [7] for the SERS spectrum of pyridine on $Ag$ with 0.2L exposure. Therefore we use the values, published in [7] for the symmetry analysis. The pyridine molecule adsorbs primary vertically, binding with the surface via nitrogen. However a part of the molecules can physisorb horizontally, or in arbitrary manner, because they are in solution or due to possible superposition of these molecules in the first layer. Then in accordance with our theory all the $Q_1 = Q_{xx}, Q_2 = Q_{yy}, Q_3 = Q_{zz}$ and $d_x, d_y, d_z$ moments may be essential for the scattering and can be the main moments for this system for various configurations. Since $Q_1, Q_2, Q_3$ and $d_z$ moments transform after the unitary irreducible representation, in accordance with the selection rules (28), the most enhanced contributions $(Q_{main} - Q_{main} - Q_{main}), (Q_{main} - Q_{main} - d_z), (Q_{main} - d_z - d_z), (d_z - d_z - d_z)$ define strongest enhancement of the bands, caused by the totally symmetric vibrations with $A_1$ symmetry (Table 2).

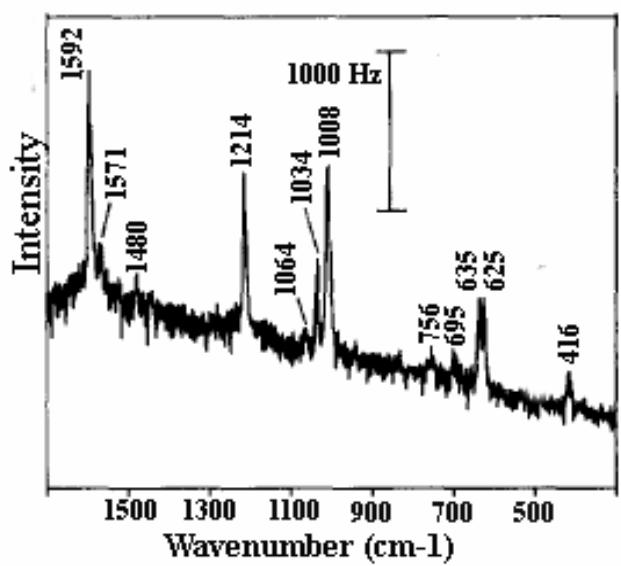

Figure 11. The SEHR spectrum of pyridine [41].

Table 2. Symmetry and wavenumbers of observable bands of the SEHR spectrum of pyridine.

| Symmetry | The mode number | Wavenumbers $(cm^{-1})$ [40] | Relative Intensity |
|---|---|---|---|
| $A_1$ | 3 | 625 | s |
| | 1 | 1008 | vs |
| | 6 | 1034 | s |
| | 8 | 1064 | vw |
| | 5 | 1214 | vs |
| | 9 | 1480 | vw |
| | 4 | 1592 | vs |
| $B_1$ | 14 | 1571 | wv |
| $B_2$ | 27 | 416 | w |
| | 26 | 695 | w |
| | 23 | 756 | w |

Appearance of the bands caused by vibrations with $B_2$ and $B_1$ symmetry is caused mainly by the $(Q_{main} - Q_{main} - d_y)$ and $(Q_{main} - Q_{main} - d_x)$ types of the scattering for pyridine lying flatly or oriented in arbitrary manner near the surface. Since the primary orientation of the adsorbed pyridine is vertical, the number of other orientations apparently is small and the intensity of the bands with $B_1$ and $B_2$ symmetry is small too. One should note, that there is a sufficiently strong line at $635 cm^{-1}$ in the experimental SEHR spectra of pyridine, published in [41, 42]. This line is absent in the pyridine spectra in [21, 22] and among the calculated wavenumbers for pyridine in [41, 42]. Therefore we do not discuss the origin of this line in the paper. Thus, after the analysis of the SEHR spectrum of pyridine we can assert that the dipole-quadrupole theory explains the SEHR spectrum of pyridine well.

## 9. Analysis of the SEHR spectrum of pyrazine

Analysis of the experimental SEHR spectrum of pyrazine obtained in [21, 22] (Figure 12) is most interesting confirmation of the dipole-quadrupole theory, since it permits to explain appearance of the strong bands of $A_g$ symmetry, which are forbidden in the usual dipole theory.

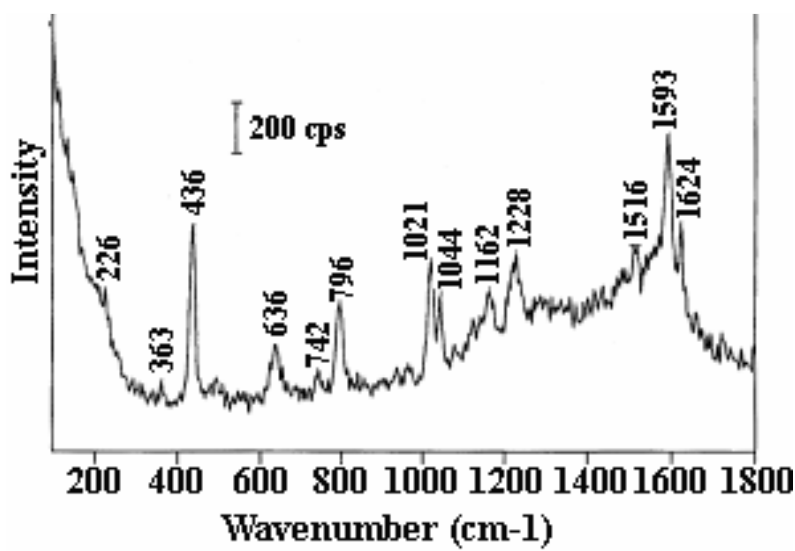

Figure 12. The SEHR spectrum of pyrazine [21]. One can see appearance of the strong bands of $A_g$ symmetry at 636, 1021, 1228, 1593 and 1624 $cm^{-1}$, arising due to the strong quadrupole light-molecule interaction.

The SEHR spectra, obtained in [21, 22] slightly differ one from another because of some difference in experimental conditions. However we have precise information about symmetry of various

SEHR bands, which is sufficient for the analysis of the spectra. Below we shall analyze the SEHR spectrum obtained in [21]. The experiments in [21] deal with the solution of pyrazine. In this case the pyrazine molecules adsorb and are oriented by three manners flatly, arbitrary and vertically binding via the nitrogen atom with the surface. Due to these possible orientations all of the $d$ moments can be essential for the scattering. In accordance with the selection rules (28) the lines with the following symmetry are observed in the SEHR spectrum (Table 3).

Table 3   Symmetry and the wavenumbers of observable bands of the SEHR spectrum of pyrazine.

| Symmetry type | Wavenumbers for SEHRS ($cm^{-1}$) | Relative intensity |
|---|---|---|
| $A_g$ | 1624 | m |
| $A_g$ | 1593 | vs |
| $A_g$ | 1228 | s |
| $A_g$ | 1021 | s |
| $A_g$ | 636 | s |
| $B_{1u}$ | 1044 | m |
| $B_{2u}$ | 796 | s |
| $B_{2u}$ | 436 | vs |
| $B_{3u}$ | 1162 | m |
| $B_{2g}$ | 1516 | w |
| $B_{3g}$ | 742 | w |

1. $A_g$ - (636, 1021, 1228, 1593 and 1624 $cm^{-1}$) caused mainly by ( $Q_{main} - Q_{main} - Q_{main}$ ) and ( $Q_{main} - d_i - d_i$ ) with $i = ( x, y, z )$ of flatly, vertically and arbitrary adsorbed pyrazine. These lines are forbidden in usual HRS and their appearance strongly proves our point of view.

2. $B_{1u}$ - 1044 $cm^{-1}$ caused mainly by the ( $Q_{main} - Q_{main} - d_z$ ) and ( $d_i - d_i - d_z$ ) ( $i = x, y, z$ ) of vertically and arbitrary oriented pyrazine respectively. The horizontally adsorbed pyrazine apparently does not determine the intensity of this band because it contains the $E_{y'}$ un-enhanced component of the electric field, which is nearly equal to zero because of the screening of the field by conductivity electrons.

3. $B_{2u} - 436$ and 796 $cm^{-1}$ caused mainly by the $(Q_{main} - Q_{main} - d_y)$ and $(d_i - d_i - d_y)$ $(i = x, y, z)$ scattering contributions of horizontally and arbitrary adsorbed molecules. The vertically adsorbed pyrazine apparently does not determine the intensity of the bands of $B_{2u}$ symmetry, since the corresponding contributions $(Q_{main} - Q_{main} - d_y)$ and $(d_z - d_z - d_y)$ include the non enhanced tangential $E_{y'}$ component of the electric field.

4. $B_{3u}$ - 1162 $cm^{-1}$ is caused mainly by the $(Q_{main} - Q_{main} - d_x)$ and $(d_z - d_z - d_x)$ scattering contributions of vertically and horizontally adsorbed pyrazine and especially by $(Q_{main} - Q_{main} - d_x)$ and $(d_i - d_i - d_x)$ $i = x, y, z$ scattering contributions of arbitrary adsorbed pyrazine. Apparently its sufficiently strong enhancement is associated mainly with the arbitrary oriented molecules. The contributions of the vertically and horizontally adsorbed pyrazine are small because they contain the un-enhanced $E_{x'}$ component of the electric field.

5. The lines of the $B_{2g}$ and $B_{3g}$ symmetry (1516 and 742 $cm^{-1}$ respectively) can be caused mainly by $(Q_{main} - d_z - d_x)$ and $(Q_{main} - d_z - d_y)$ scattering contributions of the arbitrary oriented molecules.

The absence of the lines of $A_u$ and $B_{1g}$ symmetry caused by $(d_z - d_x - d_y)$, $(Q_{main} - d_x - d_y)$ scattering contributions may be associated with the small enhancement of these lines in the spectrum due to the weak enhancement of the components of the electric field, which define the intensity of these contributions in the $x$ and $y$ directions or small number of the arbitrary oriented molecules.

Thus appearance of all the lines in the SEHR spectrum of pyrazine can be successfully explained by our theory, while appearance of the strong lines with $A_g$ symmetry strongly confirms existence of the strong quadrupole light-molecule interaction in this molecule.

## 10. Conclusion

Thus the analysis of electromagnetic fields for several models of rough surfaces and separate models of roughness points out the existence of a very large enhancement in vicinity of the tops of such models of roughness as wedges, cones, tips or spikes. For more real models of roughness the most enhancement arises in the narrow areas of the tops of the prominent places with a large curvature. Analogous areas arise between two nanowires, or two spheres, when the gap between them is very small. The limited case for these situations is the particles with sharp points like spikes. Apparently it is the most preferable configuration in order to obtain the maximum enhancement of the electric field. One should note that there will be not only the strong enhancement of the electric field, but its derivatives too. The last fact and the quantum mechanical features of the quadrupole light-molecule interaction cause the enhancement both of the dipole and quadrupole light-molecule interactions. However the most fruitful approach to the analysis of this mechanism is investigation of the SER and SEHR spectra of symmetrical molecules. Analysis of the appearance of forbidden and allowed lines in these molecules gives very precise information about the validity of this mechanism. Specific consideration of the SEHR spectra of trans-1, 2-bis (4-pyridyle) ethylene, pyridine and pyrazine permit to explain appearance of the new SEHRS bands with symmetry, which are forbidden in the pure dipole and are allowed in the dipole-quadrupole theory. Apparently the large negligence in the calculations of the wavenumbers and determination of the symmetry of the bands of trans-1, 2-bis (4-pyridyle) ethylene does not permit monosemantic determination of appearance of the bands with $A_g$ and $B_g$ symmetry in the SEHR spectra of this molecule. However numerous coincidence of the values of wavenumbers of various symmetry types in the experimental SER and SEHR spectra of this molecule permits to make a conclusion about existence of the bands of $A_g$ and $B_g$ symmetry in the SEHR spectra of this molecule, which is predicted by the dipole-quadrupole SEHRS theory. Analysis of the SEHR spectrum of pyridine confirms the possibility of explanation of its spectrum by the dipole-quadrupole theory too. However the most convincing result, which strongly supports the validity of the dipole-quadrupole theory of surface enhanced optical processes, is explanation of appearance of the forbidden bands of $A_g$ symmetry and other features of the SEHR spectra of pyrazine. More detailed mathematical exposition of this theory will be published in [43].

# 11. Appendix

## The tables of irreducible representations of some point symmetry groups

Table 4. Irreducible representations of the $C_{2v}$ and $C_{2h}$ symmetry groups.

| $C_{2v}$ | $C_1$ | $C_2(z)$ | $\sigma_v(xz)$ | $\sigma_v(yz)$ | The dipole and quadrupole moments |
|---|---|---|---|---|---|
| $A_1$ | 1 | 1 | 1 | 1 | $d_z, Q_{xx}, Q_{yy}, Q_{zz},$ |
| $A_2$ | 1 | 1 | -1 | -1 | $Q_{xy}$ |
| $B_1$ | 1 | -1 | 1 | -1 | $d_x, Q_{xz}$ |
| $B_2$ | 1 | -1 | -1 | 1 | $d_y, Q_{yz}$ |
| $C_{2h}$ | $C_1$ | $C_2(z)$ | $\sigma_h(xy)$ | I | The dipole and quadrupole moments |
| $A_g$ | 1 | 1 | 1 | 1 | $Q_{xx}, Q_{yy}, Q_{zz}, Q_{xy}$ |
| $A_u$ | 1 | 1 | -1 | -1 | $d_z,$ |
| $B_g$ | 1 | -1 | -1 | 1 | $Q_{xz}, Q_{yz}$ |
| $B_u$ | 1 | -1 | 1 | -1 | $d_x, d_y$ |

The main quadrupole moments in these groups are $Q_1 = Q_{xx}$, $Q_2 = Q_{yy}$ and $Q_3 = Q_{zz}$

Table 5. Irreducible representations of the $D_{2h}$ symmetry group.

| $D_{2h}$ | $C_1$ | $\sigma(xy)$ | $\sigma(xz)$ | $\sigma(yz)$ | $I$ | $C_2(z)$ | $C_2(y)$ | $C_2(x)$ | The dipole and quadrupole moments |
|---|---|---|---|---|---|---|---|---|---|
| $A_g$ | 1 | 1 | 1 | 1 | 1 | 1 | 1 | 1 | $Q_{xx}, Q_{yy}, Q_{zz},$ |
| $A_u$ | 1 | -1 | -1 | -1 | -1 | 1 | 1 | 1 | |
| $B_{1g}$ | 1 | 1 | -1 | -1 | 1 | 1 | -1 | -1 | $Q_{xy}$ |
| $B_{1u}$ | 1 | -1 | 1 | 1 | -1 | 1 | -1 | -1 | $d_z$ |
| $B_{2g}$ | 1 | -1 | 1 | -1 | 1 | -1 | 1 | -1 | $Q_{xz}$ |
| $B_{2u}$ | 1 | 1 | -1 | 1 | -1 | -1 | 1 | -1 | $d_y$ |
| $B_{3g}$ | 1 | -1 | -1 | 1 | 1 | -1 | -1 | 1 | $Q_{yz}$ |
| $B_{3u}$ | 1 | 1 | 1 | -1 | -1 | -1 | -1 | 1 | $d_x$ |

The main quadrupole moments in this group are $Q_1 = Q_{xx}, Q_2 = Q_{yy}$ and $Q_3 = Q_{zz}$